# POLARIZATION-DEPENDENT TRANSFORMATION OF A PARAXIAL BEAM UPON REFLECTION AND REFRACTION: A REAL-SPACE APPROACH

A.Ya. Bekshaev[*]

*I.I. Mechnikov National University, Dvorianska 2, Odessa 65082, Ukraine*

We analyze the paraxial beam transformation upon reflection and refraction at a plane boundary. In contrast to the usual approach dealing with the beam angular spectrum, we apply the continuity conditions to explicit spatial representations of the electric and magnetic fields on both sides of the boundary. It is shown that the polarization-dependent distortions of the beam trajectory (in particular, the "longitudinal" Goos–Hänchen shift and the "lateral" Imbert–Fedorov shift of the beam center of gravity) are directly connected to the incident beam longitudinal component and appear due to its transformation at the boundary.

**1. Introduction**

The Hall effect of light is a well known phenomenon demonstrating how the internal state of the light beam (polarization or inhomogeneous energy distribution) affects its trajectory (see, e.g., Refs. [1,2]). The most impressive and intensively studied manifestation is associated with lateral (out of the incidence plane) shifts of reflected and refracted beams occurring when a circularly polarized beam falls onto a plane boundary between two homogeneous media (Imbert – Fedorov shift) [3–10].

Its usual explanation is based on the incident beam representation as a superposition of partial plane waves. Each of them generates its own refracted and reflected (secondary) counterparts in accordance to the known Snell and Fresnel laws [11], so that every secondary plane wave changes the direction, amplitude and phase in respect to its prototype. Afterwards all the reflected (refracted) plane waves are put together to form the corresponding reflected (refracted) beam. Geometric phases, accepted by the circularly polarized secondary waves in compliance with their directions, lead to specific transformation of the output beam profile resulting in the Imbert–Fedorov shift [4,7–9]. Simultaneously, this shift warrants conservation of the beam angular momentum in the reflection or refraction process [1,4].

Such way of reasoning seems to be the most natural and almost automatically leads to the correct results. However, it looks rather formal. Operating in the Fourier space provides limited possibilities of employing pictorial arguments, based on the incident beam spatial structure, and role of the beam spatial parameters in the emergence and strength of the discussed effect remains obscure.

Really, consider a monochromatic paraxial beam with the wave number $k$ propagating along axis $z$ in a homogeneous medium with the refraction index $n$. The electric and magnetic fields of this beam can be represented as superpositions of two orthogonally polarized contributions denoted by subscripts $X$, $Y$ [12]:

$$\begin{Bmatrix} \mathbf{E}_X \\ \mathbf{H}_X \end{Bmatrix} = \exp(ikz) \left( \begin{Bmatrix} \mathbf{e}_x \\ n\mathbf{e}_y \end{Bmatrix} u_X + \frac{i}{k} \mathbf{e}_z \begin{Bmatrix} \partial/\partial x \\ n\partial/\partial y \end{Bmatrix} u_X \right),$$

$$\begin{Bmatrix} \mathbf{E}_Y \\ \mathbf{H}_Y \end{Bmatrix} = \exp(ikz) \left( \begin{Bmatrix} \mathbf{e}_y \\ -n\mathbf{e}_x \end{Bmatrix} u_Y + \frac{i}{k} \mathbf{e}_z \begin{Bmatrix} \partial/\partial y \\ -n\partial/\partial x \end{Bmatrix} u_Y \right) \quad (1)$$

where $u_{X,Y}(x,y,z)$ are the slowly varying complex amplitudes that satisfy the parabolic equation of paraxial optics [13]

$$i \frac{\partial u_{X,Y}}{\partial z} = -\frac{1}{2k} \nabla^2 u_{X,Y}, \quad (2)$$

---

[*] E-mail: bekshaev@onu.edu.ua

$\nabla = (\partial/\partial x, \partial/\partial y)$ is the transverse gradient. In actual fact, the paraxial field representation via Eqs. (1) and (2) goes back to the seminal work [14]. Let us focus on the longitudinal components of the field (1),

$$E_z = \frac{i}{k}\left(\frac{\partial u_X}{\partial x} + \frac{\partial u_Y}{\partial y}\right), \quad H_z = \frac{i}{k} n \left(-\frac{\partial u_Y}{\partial x} + \frac{\partial u_X}{\partial y}\right). \quad (3)$$

Within the frame of paraxial approximation, quantities (3) are small with respect to the transverse field, $E_z \sim \gamma (E_x, E_y)$. The small parameter $\gamma$ coincides with the angle of self-diffraction (beam divergence) [12–14]

$$\gamma = (k b_0)^{-1}, \quad (4)$$

$b_0$ being the characteristic size of the transverse spatial inhomogeneity of functions $u_X(x,y,z)$, $u_Y(x,y,z)$.

Let the beam be uniformly polarized,

$$u_Y(x,y,z) = \beta u_X(x,y,z), \quad (5)$$

and, additionally, axially symmetric so that in polar coordinates

$$r = \sqrt{x^2 + y^2}, \quad \phi = \arctan\left(\frac{y}{x}\right) \quad (6)$$

it is described by the azimuth-independent complex amplitude

$$u_X(x,y,z) \equiv u(x,y,z) \equiv u(r,z). \quad (7)$$

Let us inspect which features of the incident beam spatial structure are sensitive to the polarization state and how and whether can they affect the beam transformation at the plane boundary. In case of circular polarization ($\beta = \pm i$) the polarization helicity defined as [4]

$$\sigma = \frac{2 \operatorname{Im} \beta}{1 + |\beta|^2} \quad (8)$$

equals to $\sigma_\pm = \pm 1$. Obviously, the transverse beam profile completely determined by function (7) is not related to the polarization; however, its longitudinal components following from (3) with account for (5) and (6),

$$E_z = \frac{i}{k}\left(\frac{\partial}{\partial x} + i\sigma_\pm \frac{\partial}{\partial y}\right) u = \frac{i}{k} \exp(i\sigma_\pm \phi) \frac{\partial u}{\partial r},$$

$$H_z = \frac{i}{k} n \left(-i\sigma_\pm \frac{\partial}{\partial x} + \frac{\partial}{\partial y}\right) u = \sigma_\pm \frac{n}{k} \exp(i\sigma_\pm \phi) \frac{\partial u}{\partial r} \quad (9)$$

explicitly contain the polarization-dependent vortex phase factor [16]. Of course, their meaning for the energy/momentum distribution of the beam as a whole is negligible: corresponding contribution proportional to $|E_z|^2 + |H_z|^2 \sim \gamma^2$ should be discarded in the first-order paraxial approximation [14,15]. But the very existence of the components (9), that explicitly change the forms with switching the circular polarization sign, qualitatively affirms that polarization affects the beam spatial characteristics [17,18]. In fact, the "vortex" longitudinal component is the only polarization-dependent feature of the incident paraxial beam, and all the subsequent spin-sensitive effects, that may happen to the beam in the course of its propagation or transformations, inevitably "stem" from this longitudinal field. For example, relative magnitude of the $z$-components can be substantially amplified and their vortex nature comes to light due to violation of the beam paraxiality (e.g., after the beam is tightly focused [1,15,19–21]). Likewise, here we intend to show that it is the longitudinal field (9) that underlies the polarization-dependent effects accompanying the beam refraction and reflection.

### 2. General calculations

The standard scheme of the beam transformation at a plane interface is illustrated by Fig. 1 following to Ref. [22]. Plane $N$ separates two homogeneous half-spaces with refraction indices $n_I$ and $n_T$. The incident ($I$), reflected ($R$) and refracted ($T$) beams propagate along axes $z_j$ ($j = I, R, T$); with each interacting beam, its own coordinate frame is associated, all origins coinciding, axes $x_j$ lie in the plane of incidence, axis $y_j \equiv y$ is common and belongs to the boundary plane. In their own frames, the beams are described by Eqs. (1) and (2) with refraction indices $n_I$, $n_R = n_I$, $n_T$, wavenumbers $k_I$, $k_R = k_I$, $k_T$ and complex amplitudes $u^I_{X,Y}(x_I, y, z_I)$, $u^R_{X,Y}(x_R, y, z_R)$, $u^T_{X,Y}(x_T, y, z_T)$ respectively. The axes' directions are regulated by the refraction and reflection laws,

$$n_I \sin\theta_I = n_T \sin\theta_T, \quad \theta_R = \pi - \theta_I. \tag{10}$$

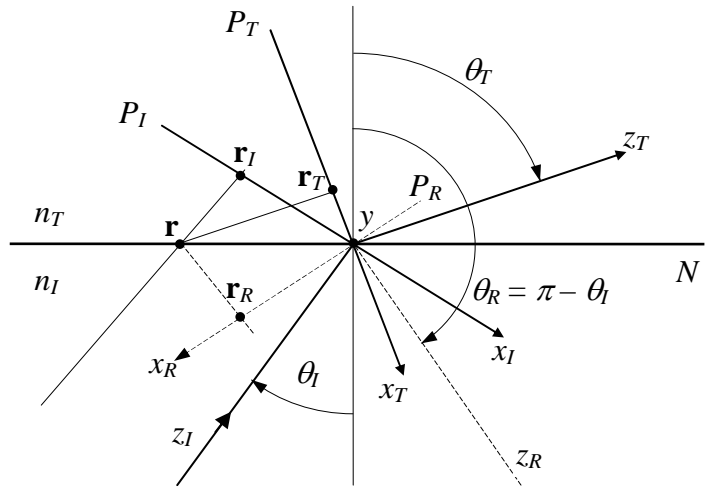

Fig. 1. Geometric conditions for the beam transformation at a plane boundary. Complex amplitudes of the incident, reflected and refracted beams are defined in reference planes $P_I$, $P_R$ and $P_T$ respectively; $\mathbf{r}_I$, $\mathbf{r}_R$ and $\mathbf{r}_T$ are the transverse coordinates of an arbitrary point $\mathbf{r}$ of the boundary measured in "own" frames associated with the incident, reflected and refracted beams. Further explanations see in text.

Relations between the three beams follow from the continuity conditions for the tangential components of the electric and magnetic fields – in fact, we operate in the usual way that gives the known Fresnel coefficients [11] but apply it to transversely confined paraxial beams rather than to plane waves. This requires some modifications of the known scheme; the main one is that any paraxial beam is naturally described only in its own cross section. Such reference cross sections are formed by planes $P_j$ defined by equations $z_j = 0$; these planes are orthogonal to axes $z_j$, and axis $y$ is their common intersection (Fig. 1). All the reference planes are not real cross sections of the corresponding beams: e.g., the incident beam does not exist in the part of plane $P_I$ above the boundary but it is suitable to characterize the incident beam by the complex amplitude distribution which would take place in plane $P_I$ if the whole half-space $z_I < 0$ were a homogeneous medium with refraction index $n_I$. Similarly, the reflected and refracted complex amplitude distributions in the reference planes $P_R$ and $P_T$ form the initial conditions for the corresponding beams that are supposed to propagate in continuous homogeneous media with indices $n_I$ and $n_T$ [1].

Our task is to link the complex amplitude distribution of the incident beam $u_{X,Y}^I(x_I, y, 0) \equiv u_{X,Y}^I(x_I, y)$, defined in the reference plane $P_I$, with functions $u_{X,Y}^R(x_R, y, 0) \equiv u_{X,Y}^R(x_R, y)$ and $u_{X,Y}^T(x_T, y, 0) \equiv u_{X,Y}^T(x_T, y)$ that characterize the secondary beams in their associated reference planes $P_R$ and $P_T$. It should be noted that boundary conditions at the interface $N$ cannot be immediately applied to these functions: at the boundary, the complex amplitudes differ from their values in the reference planes with $z_j = 0$. To find the beam fields in points where $z_j$ slightly differs from zero, one should employ Eq. (2) due to which, in the first order in $\gamma$ (first paraxial approximation),

$$u_{X,Y}^j(x_j, y, z_j) = u_{X,Y}^j(x_j, y) + \frac{iz_j}{2k_j} \nabla_j^2 u_{X,Y}^j(x_j, y), \tag{11}$$

$\nabla_j = (\partial/\partial x_j, \partial/\partial y)$. For every beam involved, plane $N$ is an oblique cross section for which

$$z_j = x_j \tan \theta_j. \tag{12}$$

Now we are in a position to write down the necessary tangential components of the electric and magnetic vectors. In accordance with (1) and (11), (12), components orthogonal to the incidence plane are associated with the $y$-oriented field vectors:

$$E_\perp^j = u_Y^j(x_j, y, x_j \tan \theta_j) \exp(ik_j x_j \tan \theta_j), \tag{13}$$

$$H_\perp^j = n_j u_X^j(x_j, y, x_j \tan \theta_j) \exp(ik_j x_j \tan \theta_j). \tag{14}$$

This representation is quite similar to its counterpart known in the case of plane-wave incidence [11]. However, the in-plane field components include contributions owing to the longitudinal field (3), and this is the main difference from the usual case:

$$E_\parallel^j = \left[ \cos\theta_j u_X^j(x_j, y, x_j \tan\theta_j) + \frac{i}{k_j} \sin\theta_j w_E^j(x_j, y) \right] \exp(ik_j x_j \tan\theta_j), \tag{15}$$

$$H_\parallel^j = n_j \left[ -\cos\theta_j u_Y^j(x_j, y, x_j \tan\theta_j) + \frac{i}{k_j} \sin\theta_j w_H^j(x_j, y) \right] \exp(ik_j x_j \tan\theta_j) \tag{16}$$

where

---

[1] Such a choice of reference planes is convenient in calculations and makes no trouble for interpretations because any of the three beams can only be observed at a certain distance from the boundary, where it is situated entirely within the corresponding homogeneous medium.

$$w_E^j(x_j, y) = \frac{\partial u_X^j(x_j, y)}{\partial x_j} + \frac{\partial u_Y^j(x_j, y)}{\partial y}, \tag{17}$$

$$w_H^j(x_j, y) = \frac{\partial u_X^j(x_j, y)}{\partial y} - \frac{\partial u_Y^j(x_j, y)}{\partial x_j}. \tag{18}$$

Formally, Eqs. (17) and (18) must have contained the complex amplitude derivatives taken at points of the boundary, i.e. with corrections for the propagation distances between the reference plane and the boundary, see Eq. (11). However, the "longitudinal" terms of (15) and (16) 'per se' are of the first order of smallness due to $k_j$ in denominators, so the second terms of (11) would provide the second-order corrections and are thus neglected in the derivatives of (17) and (18).

Then, expressions (13) – (18) should be substituted into the usual continuity conditions [11]

$$E_\perp^I + E_\perp^R = E_\perp^T, \quad H_\perp^I + H_\perp^R = H_\perp^T; \tag{19}$$

$$E_\parallel^I + E_\parallel^R = E_\parallel^T, \quad H_\parallel^I + H_\parallel^R = H_\parallel^T \tag{20}$$

which, together with (13) – (18), yield four equations to determine the four unknowns $u_{X,Y}^R(x_R, y)$ and $u_{X,Y}^T(x_T, y)$ for given complex amplitudes of the incident beam $u_{X,Y}^I(x_I, y)$. Note that in Eqs. (13) – (18), spatial functions pertaining to the *j*-th beam are expressed in their own coordinates; in fact, they relate to the same points of the boundary plane (see relations between $\mathbf{r}_I$, $\mathbf{r}_R$ and $\mathbf{r}_T$ in Fig. 1) and the *x*-coordinates in different terms of (13) – (18) are linked by equations

$$x_R = -x_I, \quad x_T \cos\theta_I = x_I \cos\theta_T. \tag{21}$$

In the "differential" terms of (17), (18), substitution (21), if necessary, should be made after the differentiation with respect to the "own" $x_j$ is performed.

After substitution of expressions (13) – (16) into (19), (20), we obtain a complete set of conditions determining the vector beam transformation at a plane boundary. They can be solved by successive approximations. First, in view of Eqs. (10) and (21), all the exponential multipliers in (13) – (16) appear to be identical and may be omitted. Then, approximate solution to Eqs. (19), (20) is sought in the form

$$u_{X,Y}^j(x_j, y) = \tilde{u}_{X,Y}^j(x_j, y) + \delta u_{X,Y}^j(x_j, y) \quad (j = R, T) \tag{22}$$

where $\tilde{u}_{X,Y}^j(x_j, y)$ allows for the zero-order terms and $\delta u_{X,Y}^j(x_j, y)$ takes into account corrections of the order $\gamma$ associated with the beams' transformation on passages between the reference planes and the boundary (second terms in (11)) and with the longitudinal field contributions (second terms in brackets of (15), (16)). For the zero-order terms one easily obtains obvious relations

$$\tilde{u}_X^T(x_T, y) = T_\parallel u_X^I(x_I, y), \quad \tilde{u}_Y^T(x_T, y) = T_\perp u_Y^I(x_I, y), \tag{23}$$

$$\tilde{u}_X^R(x_R, y) = R_\parallel u_X^I(x_I, y), \quad \tilde{u}_Y^R(x_R, y) = R_\perp u_Y^I(x_I, y) \tag{24}$$

where

$$T_\parallel = \frac{2n_I \cos\theta_I}{n_T \cos\theta_I + n_I \cos\theta_T}, \quad T_\perp = \frac{2n_I \cos\theta_I}{n_I \cos\theta_I + n_T \cos\theta_T}, \tag{25}$$

$$R_\parallel = \frac{n_T \cos\theta_I - n_I \cos\theta_T}{n_T \cos\theta_I + n_I \cos\theta_T}, \quad R_\perp = \frac{n_I \cos\theta_I - n_T \cos\theta_T}{n_I \cos\theta_I + n_T \cos\theta_T} \tag{26}$$

are the usual Fresnel transmission and reflection coefficients [11]. Note that in Eqs. (23), (24) arguments $x_I$, $x_R$ and $x_T$ are connected by Eqs. (21).

For calculation of the corrections $\delta u^j_{X,Y}(x_j, y)$ of Eq. (22), results of zero approximation (23), (24) are substituted into the terms of (13) – (16) that are proportional to $k_j^{-1}$. Then, after simple but tedious algebra, involving relations (23) – (26), (21) and the derivative correspondences

$$\frac{\partial}{\partial x_R} = -\frac{\partial}{\partial x_I}, \quad \frac{\partial}{\partial x_T} = \frac{\cos\theta_I}{\cos\theta_T}\frac{\partial}{\partial x_I},$$

one arrives at the final results which, in view of their importance, are presented in explicit form:

$$\delta u^T_X(x_T, y) = \frac{i}{2k_I} x_I \tan\theta_I\, T_\parallel \nabla^2_I u^I_X(x_I, y) - \frac{i}{2k_T} x_T \tan\theta_T\, T_\parallel \nabla^2_T u^I_X(x_I, y)$$
$$+ \frac{i}{2k_I}\tan\theta_I \frac{n_I}{n_T} T_\parallel \left[ T_\parallel C_\parallel \frac{\partial u^I_X(x_I, y)}{\partial x_I} + T_\perp C_\perp \frac{\partial u^I_Y(x_I, y)}{\partial y} \right], \quad (27)$$

$$\delta u^R_X(x_R, y) = \frac{i}{2k_I}\tan\theta_I\, T_\parallel \left[ T_\parallel C_\parallel \frac{\partial u^I_X(x_I, y)}{\partial x_I} + T_\perp C_\perp \frac{\partial u^I_Y(x_I, y)}{\partial y} \right], \quad (28)$$

$$\delta u^T_Y(x_T, y) = \frac{i}{2k_I} x_I \tan\theta_I\, T_\perp \nabla^2_I u^I_Y(x_I, y) - \frac{i}{2k_T} x_T \tan\theta_T\, T_\perp \nabla^2_T u^I_Y(x_I, y)$$
$$+ \frac{i}{2k_I}\tan\theta_I\, T_\perp \left[ T_\perp C_\parallel \frac{\partial u^I_Y(x_I, y)}{\partial x_I} - T_\parallel C_\perp \frac{\partial u^I_X(x_I, y)}{\partial y} \right], \quad (29)$$

$$\delta u^R_Y(x_R, y) = \frac{i}{2k_I}\tan\theta_I\, T_\perp \left[ T_\perp C_\parallel \frac{\partial u^I_Y(x_I, y)}{\partial x_I} - T_\parallel C_\perp \frac{\partial u^I_X(x_I, y)}{\partial y} \right]. \quad (30)$$

Here

$$C_\parallel = \frac{n_T \cos\theta_T}{n_I \cos\theta_I} - \frac{n_I \cos\theta_I}{n_T \cos\theta_T}, \quad C_\perp = \frac{n_T}{n_I} - \frac{n_I}{n_T}, \quad (31)$$

arguments $x_I$, $x_R$ and $x_T$ are still related by Eqs. (21).

Eqs. (27) – (31) represent a vector generalization of the scalar formulae obtained before [22,23]. The vector nature of the optical field manifests itself by the fact that Eqs. (27), (28) determining $\delta u^{R,T}_X$, contain not only the in-plane incident field $u^I_X$ but also derivatives of the orthogonal component $u^I_Y$, and Eqs. (29), (30) for $\delta u^{R,T}_Y$ contain contributions of $u^I_X$. Interestingly, these "vector" modifications involve derivatives with respect to the "off-plane" coordinate $y$. Without them, Eqs. (27) – (30) can be reduced to relations derived in Refs. [22,23] for a boundary with diffraction grating if the grating period tends to infinity ("smooth" interface). Indeed, due to relations

$$\frac{1}{2}\tan\theta_I T^2_\parallel C_\parallel = -\frac{n_T}{n_I}\frac{dT_\parallel}{d\theta_I} = -\frac{dR_\parallel}{d\theta_I}, \quad \frac{1}{2}\tan\theta_I T^2_\perp C_\parallel = -\frac{dT_\perp}{d\theta_I} = -\frac{dR_\perp}{d\theta_I}, \quad (32)$$

the terms of Eqs. (27) – (30), proportional to derivatives with respect to $x_I$, can be expressed via the angular derivatives of the corresponding transformation coefficients, as is suggested by the angular-spectrum-based reasoning [4,22][2]. Importantly, no special requirements were imposed so far relating the real or complex values of $n_I$ and $n_T$, and expressions (23) – (31), though presented formally for a dielectric interface, are applicable in cases where one or both contacting media are conductive.

---

[2] In Refs. [22,23] the plane wave transformation at the boundary is characterized by "amplitude efficiencies" $\tau(\theta_I) = \sqrt{n_T \cos\theta_T / n_I \cos\theta_I}\, T_{\parallel,\perp}(\theta_I)$ rather than by the "pure" transmission coefficients $T_{\parallel,\perp}(\theta_I)$; this results in additional summands in the expression of $d\tau/d\theta_I$ as compared to $dT_{\parallel,\perp}/d\theta_I$, which apparently caused additional terms with $x$-derivatives in formulae for the secondary beam deformations of [22,23].

The reflected/refracted beam spatial configuration described by Eqs. (23), (24) and (27) – (30) completely coincides with what follows from the traditional approach based on the plane-wave expansion [4,9] in the first order in $\gamma$. However, the presented way of reasoning discloses some new important aspects of the beam transformation. Note that all terms with first derivatives in (27) – (30) appear due to the longitudinal component of the incident field. It is important to emphasize that the *longitudinal* component of the incident beam contributes to formation of the *transverse* components of the reflected and refracted beams. Therefore, while in the incident field the small corrections of the order $\gamma$ were orthogonal to the "main" transverse field and did not disturb the beam energy distribution in the first paraxial order (see notes below Eq. (9)), in the secondary (reflected/refracted) beams, the first-order perturbation "penetrates" into the transverse components and can cause quite perceptible distortions of their spatial profile. The terms of Eqs. (27) – (30) proportional to $\partial/\partial x_I$ describe distortions "oriented" parallel to the incidence plane; generally, they are responsible for the "in-plane" beam shift known as the Goos–Hänchen shift [4,9,10], the best observable in conditions of the total reflection. The "out-of-plane" Imbert–Fedorov shift [3,4], orthogonal to the incidence plane, is associated with terms containing $\partial/\partial y$. Our present consideration shows that both effects can be treated as immediate consequences of the non-transversality of the incident beam field and directly originate from its longitudinal component.

### 3. The secondary beam shift

To demonstrate equivalence of the presented approach and the previously developed models [3–9], let us determine the lateral shift of the refracted beam described by Eqs. (23), (27), (29). In general, the center of gravity coordinates for the *j*-th beam are defined as

$$\begin{Bmatrix} x_0^j \\ y_0^j \end{Bmatrix} = \frac{\int \begin{Bmatrix} x \\ y \end{Bmatrix} \left( \left| u_X^j \right|^2 + \left| u_Y^j \right|^2 \right) dx_j \, dy}{\int \left( \left| u_X^j \right|^2 + \left| u_Y^j \right|^2 \right) dx_j \, dy} \tag{33}$$

(integration over the whole beam cross section is implied). Terms $x_0^{R,T}$ correspond to the Goos – Hänchen shift; the center of gravity displacements orthogonal to the incidence plane (Imbert – Fedorov shifts) are associated with $y_0^{R,T}$. Hence, in application to the refracted beam, we have

$$y_0^T = \frac{\int y \left( \left| u_X^T \right|^2 + \left| u_Y^T \right|^2 \right) dx_T \, dy}{\int \left( \left| u_X^T \right|^2 + \left| u_Y^T \right|^2 \right) dx_T \, dy} \approx \frac{\int y \left( T_\parallel^2 \left| u_X^I \right|^2 + T_\perp^2 \left| u_Y^I \right|^2 \right) dx_I \, dy}{\int \left( T_\parallel^2 \left| u_X^I \right|^2 + T_\perp^2 \left| u_Y^I \right|^2 \right) dx_I \, dy}$$

$$+ \frac{\int y \left\{ T_\parallel \left[ \left( u_X^I \right)^* \delta u_X^T + u_X^I \left( \delta u_X^T \right)^* \right] + T_\perp \left[ \left( u_Y^I \right)^* \delta u_Y^T + u_Y^I \left( \delta u_Y^T \right)^* \right] \right\} dx_I \, dy}{\int \left( T_\parallel^2 \left| u_X^I \right|^2 + T_\perp^2 \left| u_Y^I \right|^2 \right) dx_I \, dy} \tag{34}$$

(asterisk, as usual, denotes the complex conjugate, and both media are supposed dielectric so that transmission coefficients (25), (26) are real). If the incident beam possesses circular symmetry (e.g., is Gaussian), the first summand in the right-hand side of (34) vanishes, while in the second summand (second line of (34)), only terms containing derivatives with respect to $y$ can give non-zero contributions. Consequently, expression in figure brackets of (34) may be replaced by

$$\frac{i}{2k_I} \tan \theta_I \, T_\perp T_\parallel C_\perp \left\{ \frac{n_I}{n_T} T_\parallel \left[ \left( u_X^I \right)^* \frac{\partial u_Y^I}{\partial y} - u_X^I \frac{\partial \left( u_Y^I \right)^*}{\partial y} \right] - T_\perp \left[ \left( u_Y^I \right)^* \frac{\partial u_X^I}{\partial y} - u_Y^I \frac{\partial \left( u_X^I \right)^*}{\partial y} \right] \right\}. \tag{35}$$

This expression includes only parameters of the incident beam profile and we may omit the super- and subscripts "$I$" in further analysis. Now suppose the incident beam to be homogeneously polarized; together with assumed circular symmetry this means that Eqs. (5) and (7) are true. If, additionally, the complex amplitude function $u(x,y)$ is real (e.g., the beam waist plane coincides with the reference plane $P_I$, Fig. 1), denominator of Eq. (34) acquires the form $\left(T_\parallel^2 + |\beta|^2 T_\perp^2\right)\int u^2 dx dy$ and expression in figure brackets of (35) reduces to

$$2i\,\text{Im}\,\beta\left(\frac{n_I}{n_T}T_\parallel + T_\perp\right)u\frac{\partial u}{\partial y}.$$

After this is substituted into (34), the whole numerator of (34) can be transformed by using the integral relation

$$\int y u \frac{\partial u}{\partial y} dx\,dy = -\frac{1}{2}\int u^2 dx\,dy$$

(valid for any function $u(x,y)$ that tends to zero quickly enough at the transverse infinity), which ultimately yields

$$y_0^T = \frac{\tan\theta_I}{2k_I}\frac{\text{Im}\,\beta}{T_\parallel^2 + |\beta|^2 T_\perp^2} C_\perp T_\parallel T_\perp \left(\frac{n_I}{n_T}T_\parallel + T_\perp\right). \tag{36}$$

For the reflected beam, lateral shift can be found similarly,

$$y_0^R = \frac{\tan\theta_I}{2k_I}\frac{\text{Im}\,\beta}{R_\parallel^2 + |\beta|^2 R_\perp^2} C_\perp T_\parallel T_\perp \left(R_\parallel + R_\perp\right). \tag{37}$$

With the help of relations (10), (25), (26) and (31), one can make sure that these results are identical to more familiar relations obtained via the known theory of the spin Hall effects; e.g., by means of the easily verified equality $\tan\theta_I C_\perp T_\parallel T_\perp = -2\cot\theta_I\left(R_\parallel + R_\perp\right)$, Eqs. (36) and (37) can be reduced to Eq. (58) of Ref. [4].

The angular Goos – Hänchen and Imbert – Fedorov shifts [3,4] can be found by a similar procedure which, instead of (33), starts with expression for the mean tilt of the secondary beam trajectory [15,22,27]

$$\begin{pmatrix} p_{x0}^j \\ p_{y0}^j \end{pmatrix} = -\frac{i}{k_j}\frac{\int\left[\left(u_X^j\right)^*\nabla_j u_X^j + \left(u_Y^j\right)^*\nabla_j u_Y^j\right]dx_j\,dy}{\int\left(|u_X^j|^2 + |u_Y^j|^2\right)dx_j\,dy} \quad (j=R,T)$$

where $p_{x0}^j = n_j\dfrac{dx_0^j}{dz_j}$ and $p_{y0}^j = n_j\dfrac{dy_0^j}{dz_j}$.

### 4. Conclusion

We have considered transformation of a paraxial beam upon reflection and/or refraction at a plane boundary between two homogeneous media. In contrast to the usual approaches based on the field representation via the plane-wave spectrum (in the momentum space) [3–10], our consideration relies upon the real-space arguments. It starts with representation of a vector paraxial beam via complex amplitudes of orthogonally polarized components, valid in the first order of paraxial approximation (first order in the divergence angle $\gamma$). The main feature of this representation is explicit presence of the longitudinal field component. Analysis of the beam transformation at the boundary is performed by direct generalization of the usual procedure developed for plane waves and commonly used for obtaining the Fresnel laws for the refraction and reflection coefficients [11]. Deviations from the geometric picture of the beam transformation are considered as immediate consequences of the first-order corrections to the incident beam spatial structure. In particular, it is the longitudinal electric and magnetic fields of the incident beam that give rise to such known post-geometric effects as the polarization-dependent Goos – Hänchen and Imbert – Fedorov shifts. Especially, the Imbert – Fedorov shift of the beam center of gravity can be treated as a manifestation of the polarization-sensitive vortex structure of the incident beam longitudinal component. In the course of reflection or refraction, this vortex structure partly "penetrates" into transverse components of the reflected and refracted beams and causes the observable spin-dependent distortion of their spatial profiles. This reveals relationship between the Imbert – Fedorov shift at a plane boundary and the lateral shift of a focal spot [17] – another spin Hall effect manifestation that also appears due to the longitudinal field modification caused by switching the sign of circular polarization. In both phenomena, the longitudinal field of the initial paraxial beam appears as a crucial element of the mechanism by which the spin-orbit interaction is realized. In conjunction with the decisive role that the longitudinal component plays in the spin-to-orbit angular momentum conversion upon the light beam focusing [19–21] or scattering [24], this leads to suggestion of the special importance of the longitudinal field in other spin-orbit phenomena involving paraxial beams, which should be clarified in further research.

The "real-space" approach to the beam transformation at a plane boundary appears to be completely equivalent to the usual plane-wave-spectrum method of analysis. However, the presented way of operation, essentially based on pictorial geometric arguments, may be useful in some methodical aspects. Its final results (Eqs. (27) – (30) with known auxiliary relations (23) and (24)) are valid for arbitrary paraxial beam and describe, from the common point of view, not only spin-dependent effects but also the beam transformations owing to its internal spatial structure (e.g., orbital Hall effect [15,25,26]). We hope the method of this paper will be useful in further studies of the optical spin Hall phenomena, including elucidation of the role and manifestation of the internal energy flows [15].

**Acknowledgements**

The author is grateful to K. Bliokh for helpful discussion.